\documentclass[11pt,twoside]{article}
\usepackage{asp2010}
\resetcounters

\begin{document}

\title{Precision Calibration via Artificial Light Sources Above the Atmosphere}
\author{Justin E.~Albert$^1$, Maxwell H.~Fagin$^2$, Yorke J.~Brown$^{2,3}$, Christopher W. Stubbs$^3$, Nikita A.~Kuklev$^1$, and Alexander J.~Conley$^4$
\affil{$^1$Department of Physics \& Astronomy, University of Victoria, Victoria BC V8P5C2 Canada}
\affil{$^2$Department of Physics \& Astronomy, Dartmouth College, Hanover NH 03755 USA}
\affil{$^3$Department of Physics \& Department of Astronomy, Harvard University, 17 Oxford Street Cambridge MA 02138 USA}
\affil{$^4$Center for Astrophysics and Space Astronomy, University of Colorado, Boulder CO 80304 USA}}

\begin{abstract}
Deeper understanding of the properties of dark energy via SNIa surveys, and to a large extent other methods as well, will require
unprecedented photometric precision.  Laboratory and solar photometry and radiometry regularly achieve precisions on the 
order of parts in ten thousand, but photometric calibration for non-solar astronomy presently remains stuck at the percent or 
greater level.  We discuss our project to erase this discrepancy, and our steps toward achieving laboratory-level photometric
precision for surveys late this decade.  In particular, we show near-field observations of the balloon-borne light source we
are presently testing, in addition to previous work with a calibrated laser source presently in low-Earth orbit.  Our technique
is additionally applicable to microwave astronomy.  Observation of gravitational waves in the polarized CMB will similarly
require unprecedented polarimetric and radiometric precision, and we briefly discuss our plans for a calibrated microwave
source above the atmosphere as well.
\end{abstract}

\section{INTRODUCTION}

\begin{figure}[!t]
\begin{center}
\includegraphics[angle=0,width=6.5cm]{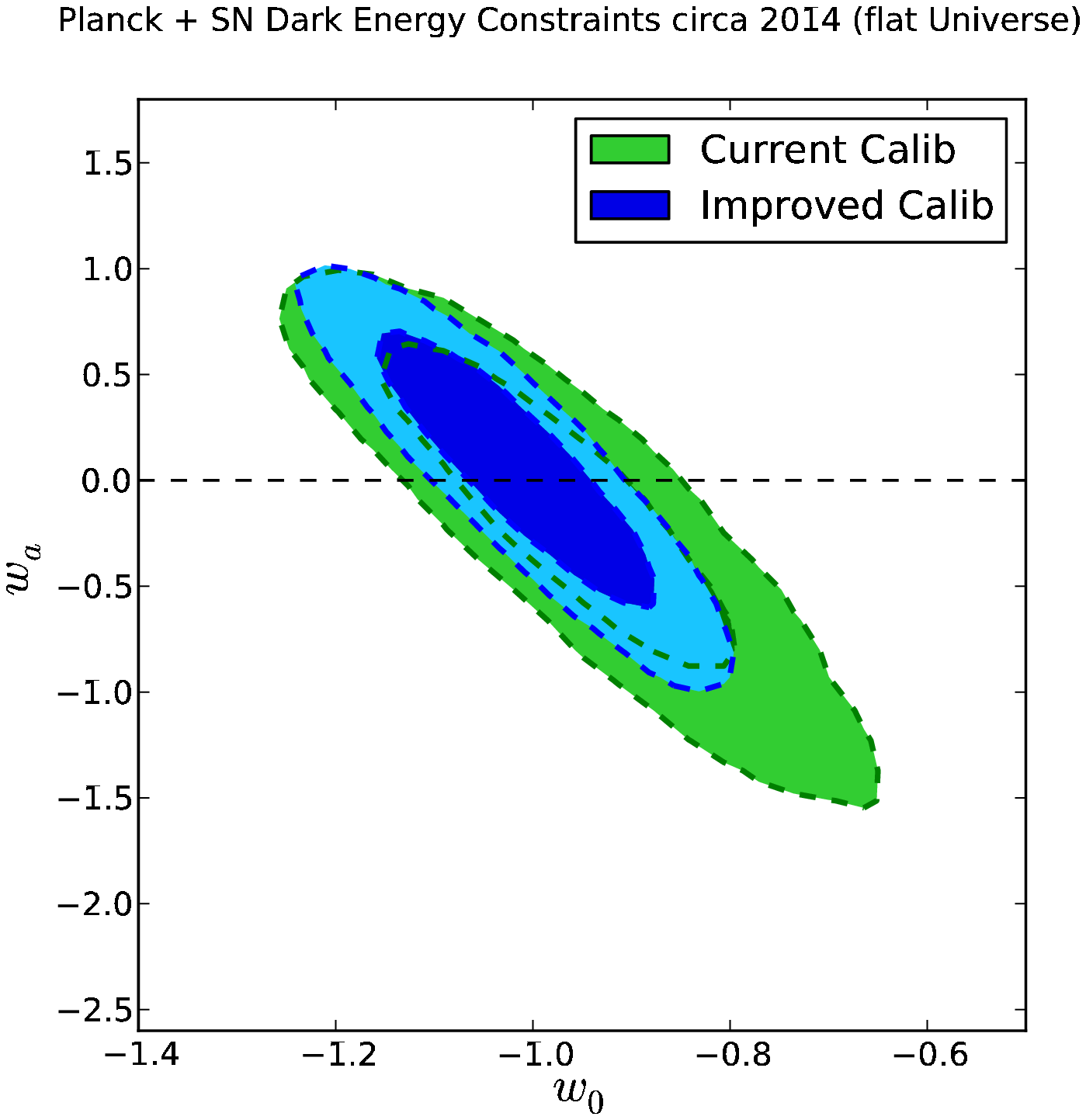}
\hfill
\includegraphics[angle=0,width=6.5cm]{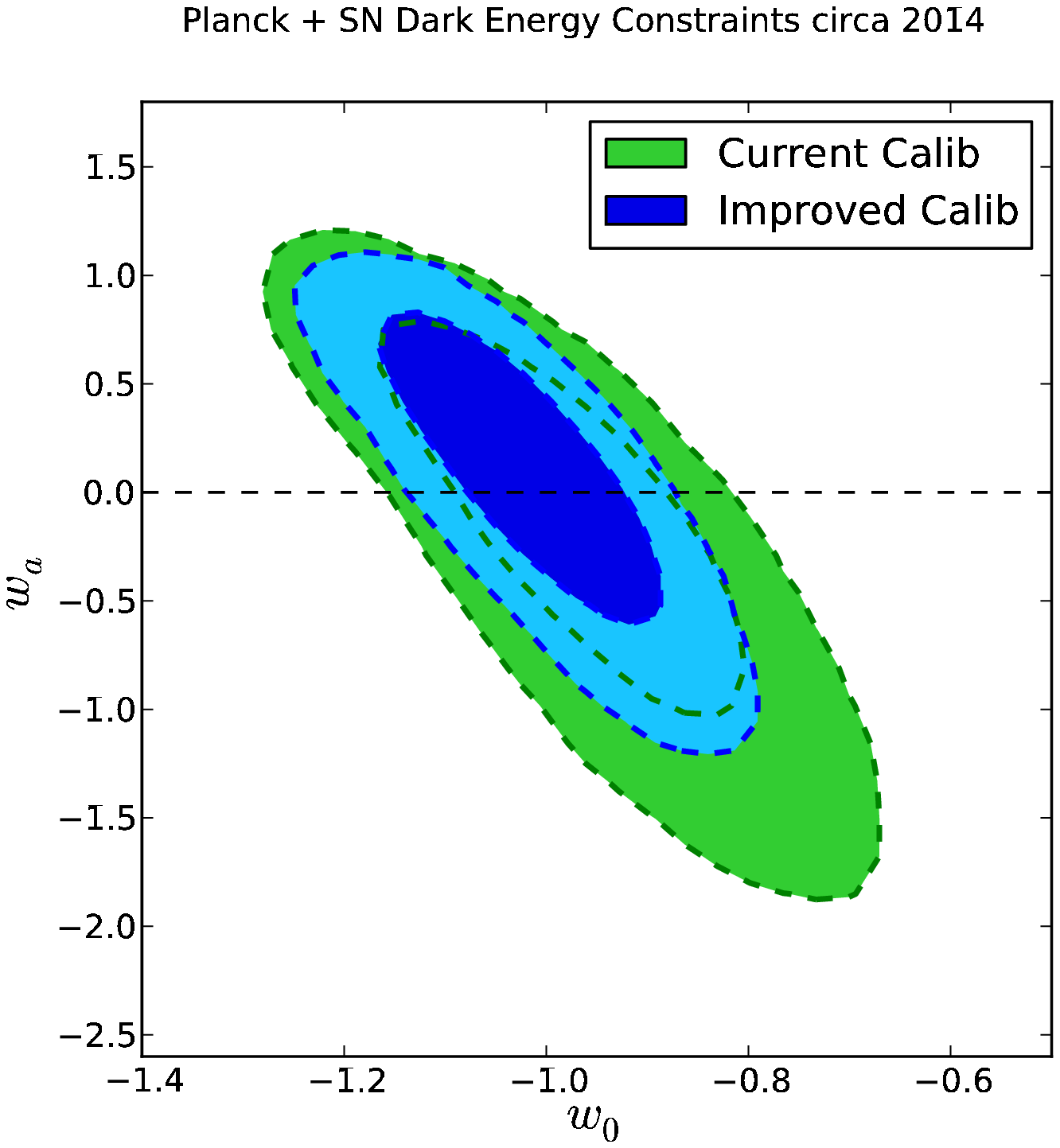}
\caption{
Constraints on the dark energy equation of state parameters w$_0$ ($\equiv p/\rho$ of the universe at the present day), and
w$_a \equiv$ dw/d$a$ (also at present, where $a$ is the scale factor of the universe), for all current and upcoming SNIa dark
energy projects (plus the Planck CMB mission) with the new calibration provided by our calibration program, vs.~with all current
means of calibration.  The left plot shows the uncertainties if one artificially constraints the universe to be flat, and the right 
plot shows the uncertainties with flatness constraint removed.  Dashed lines show 68\% and 95\% confidence intervals for each case.
The calibration will improve the dark energy ``figure of merit''~\citep{alb06} by a factor of 2.4, with additional improvement beyond
2014.  
}
\label{fig:Alexplots}
\end{center}
\end{figure}

The presence of artificial flux standards above the Earth's atmosphere may provide 
significant reduction of photometric uncertainties for
measurements that depend on such calibration.  
The combined effect of atmospheric and instrumental extinction in the visible and near-infrared is presently the source of 
the largest uncertainty on the properties and 
amount of dark energy (see Fig.~\ref{fig:Alexplots}).  
Man-made visible and NIR light sources 
can be measured to a precision of up to 100 times
better than standard stellar sources, as presently shown by
both laboratory and solar irradiance measurements,
allowing for vast reduction of astronomical and cosmological uncertainties due to photometry.

Separately, in the microwave spectrum, polarization patterns in the cosmic microwave 
background (CMB) encode a vast amount of other information (beyond dark 
energy) about the early universe, ranging from the amplitude of 
primordial gravitational waves and the energy scale of cosmic inflation, 
to the gravitational lensing potential integrated over cosmic history. 
The interpretation of these signals is entirely contingent on an 
accurate calibration of polarized instrumental sensitivity. As yet, no 
well-calibrated, polarized, celestial microwave sources exist, and in 
the coming generation of microwave telescopes, this deficit may become
a limiting factor in our ability to measure and understand the 
polarized microwave sky. 
Current calibration solutions rely on relatively nearby ground-based 
sources, requiring refocusing and special low-sensitivity detectors to 
handle the near-field source and additional atmospheric loading. The 
cleanest and simplest calibration solution would solve both of these 
issues, by lofting a source into the far field, at distances of order 
tens of kilometres on large microwave telescopes.

Prior to the second half of the 20th century, the only sources of light above the Earth's atmosphere were natural in origin: stars, and 
reflected light from planets, moons, comets, etc.  Natural sources have of course served extremely well in astronomy: through 
understanding the physical processes governing stellar evolution, we are now able to fairly precisely understand the spectra of stars used 
as calibration sources [\textit{e.g.} \citet{boh00}].  Nevertheless, in all stars the vast bulk of material, and the thermonuclear 
processes that themselves 
provide the light, lie beyond our sight below the surface of the star.  Superb models of stellar structure are available, but 
uncertainties of many types always remain.

Since the launching of the first high-altitude balloons and satellites, a separate class of potential light sources in 
space and in near-space has become available. 
Observable light from most satellites is primarily due to direct solar reflection, or reflection from Earth's albedo.  While 
providing a convenient method of observing satellites, this light is typically unsuitable for use as a calibrated light source due 
to large uncertainties in the reflectivity (and, to a lesser extent, the precise orientation and reflective area) of satellites' 
surfaces.\footnote{Reflected solar light has, however, been successfully used as an absolute infrared calibration source by the Midcourse 
Space Experiment (MSX), using 2 cm diameter black-coated spheres ejected from the MSX satellite, whose infrared emission was 
monitored by the instruments aboard MSX~\citep{pri04}.  This technique proved highly effective for the MSX infrared calibration; 
however, the technique is not easily applicable to measuring extinction of visible light in the atmosphere.}

However, balloon- and satellite-based calibration sources for ground-based telescopes are by no means technically prohibitive.
As an example, a standard household 25-watt tungsten filament lightbulb (which typically have a temperature of the order of
3000 K and usually produce approximately 1 watt of visible light between 390 and 780 nm) which
radiates light equally in all directions from a 700 km low Earth orbit has an equivalent brightness to a 12.5-magnitude star
(in the AB system, although for this approximate value the system makes little difference).  In
general, the apparent magnitude of an orbiting lamp at a typical incandescent temperature which radiates isotropically is 
approximately given by
\begin{equation}
m \approx -5.0 \log_{10} \left ( \frac{ \left ( \ln \left ( \frac {P}{\rm{1\;watt}} \right ) \right )^3 }{h} \right ) + 5.9,  
\end{equation}
where $P$ is the power of the lamp in watts, and $h$ is the height of the orbit in kilometers.
The systematic 
uncertainty on the radiance of an optimally-designed above-atmosphere lamp, where cost is no object, would be dominated by the precision of
radiometric monitoring technology, and be in the range of approximately 200 parts per million if the best presently available 
radiometric technology is used.

An alternative to an isotropic or near-isotropic lamp would be a laser source, with beam pointed at the observer (with a small 
moveable mirror, for example).  Divergences of laser beams are typically on the order of a milliradian (which can be reduced to 
microradians with a beam expander) so much less output power than a lamp would be required for a laser beam to mimic the brightness 
of a typical star. 
The apparent magnitude of an orbiting laser with Gaussian beam divergence 
pointed directly at a ground-based telescope, is given by
\begin{equation}
m \approx -2.5 \log_{10}\left ( \frac{P}{h^2 d^2} \right ) - 20.1,
\end{equation}
where $P$ is the laser power in milliwatts, $h$ is the
height of the orbit in kilometers, and $d$ is the RMS divergence of the laser beam in milliradians,
under the assumption that the aperture of the telescope is small compared with the RMS width of the beam at the ground, $hd$.
The RMS divergence would be the combination of the divergence at the source, and the divergence due to the atmosphere.  In
clear conditions, total atmospheric divergence in a vertical path is at the level of approximately 5 microradians~\citep{tat61}, 
and this of course only
acts on the last fraction of the laser path that is within the atmosphere, so as long as the source divergence is
significantly larger than this, atmospheric divergence would be negligible. 
The uncertainty on the apparent magnitude of an orbiting laser stemming from uncertainties in the radiometrically-monitored laser 
power would likely be limited by the precision of current radiometer technology.  Modern electrical substitution 
radiometers can achieve a precision of approximately 100 parts per million when aperture uncertainties can be neglected, as in the
case of laser radiometry~\citep{kop05}.  
Uncertainties on the magnitude due to 
uncertainty in the pointing and beam profile would potentially be limited by the size of the array of outboard telescopes for 
monitoring the laser spot, and by calibration differences between the individual telecopes in the array and with the main central 
telescope.  The latter could clearly be minimized by a ground system for ensuring the relative calibration of the outboard 
telescopes and main telescope are all consistent.  

The uncertainties considered above assume that the exposure time is long compared with the coherence time of the atmosphere.  With 
short exposures --- or in the case of a laser that either quickly sweeps past, or is pulsed --- atmospheric scintillation can play a 
major role in uncertainty in apparent magnitude of an above-atmosphere source.  A typical timescale for a CW laser with 1 
milliradian divergence in low Earth orbit to sweep past is tens of milliseconds, which is of the same order as characteristic 
timescales of atmospheric scintillation, and the typical timescale of single laser pulses is nanoseconds, much shorter than 
scintillation timescales, thus one cannot assume that such effects can be time-averaged over.  In idealized conditions, for small 
apertures $D < \sim 5$ cm and sub-millisecond integration times, the relative standard deviation in intensity $\sigma_I \equiv 
\Delta I / \langle I \rangle$, where $\Delta I$ is the root-mean-square value of $I$, is given by the square root of
\begin{equation}
\sigma_I^2 = 19.12 \lambda^{-7/6} \int_0^\infty C_n^2(h)h^{5/6}dh,
\end{equation}
where $\lambda$ is optical wavelength (in meters), $C_n^2(h)$ is known as the refractive-index structure coefficient, and $h$ is 
altitude (in meters) \citep{tat61}.  Large apertures $D > \sim 50$ cm have a relative standard deviation in intensity given by the square root of
\begin{equation}
\sigma_I^2 = 29.48 D^{-7/3} \int_0^\infty C_n^2(h)h^{2}dh
\end{equation}
\citep{tat61}. The values and functional form of $C_n^2(h)$ are entirely dependent on the particular atmospheric conditions at the 
time of observation, however a relatively typical profile is given by the Hufnagel-Valley form:
\begin{eqnarray}
\!\!\! C_n^2(h) \! & \!\!\! = \!\!\! & \! 5.94 \times 10^{-53}(v/27)^2 h^{10}e^{-h/1000} + \nonumber\\
                \! &                 & \! 2.7 \times 10^{-16} e^{-h/1500} + Ae^{-h/100},
\end{eqnarray}
where $A$ and $v$ are free parameters \citep{huf74}.  Commonly-used values for the $A$ and $v$ parameters, which represent the 
strength of turbulence near ground level and the high-altitude wind speed respectively, are $A = 1.7 \times 10^{-14}$ m$^{-2/3}$ 
and $v = 21$ m/s \citep{rog96}. Using these particular values, for a small aperture, the relative standard deviation $\sigma_I$ would be 
expected to be 0.466 for 532 nm light, which is not far off experimental scintillation values for a clear night at a 
typical location [\textit{e.g.}~\citet{jak78}]. For a single small camera, this is an extremely large uncertainty.  Other than by 
increasing integration time (which is not possible with a pulsed laser) or by significantly increasing the camera aperture, the 
only way to reduce this uncertainty is to increase the number of cameras.  With $N$ cameras performing an observation, which are 
spaced further apart than the coherence length of atmospheric turbulence (typically 5 to 50 cm), the uncertainty from scintillation 
can be reduced by a factor $\sqrt{N}$ (for large $N$).

The considerations above are necessarily both speculative and 
approximate.  However, at present there is an actual laser in low Earth orbit, visible with both equipment and with the naked eye, 
and analysis of ground-based observational data of the laser spot can be used for comparisons with the above, as well as for 
development of and predictions for potential future satellite-based photometric calibration sources of ground telescopes.

\begin{figure*}[p]
\vspace*{-8mm}
\begin{center}
\includegraphics[angle=0,width=6.5cm]{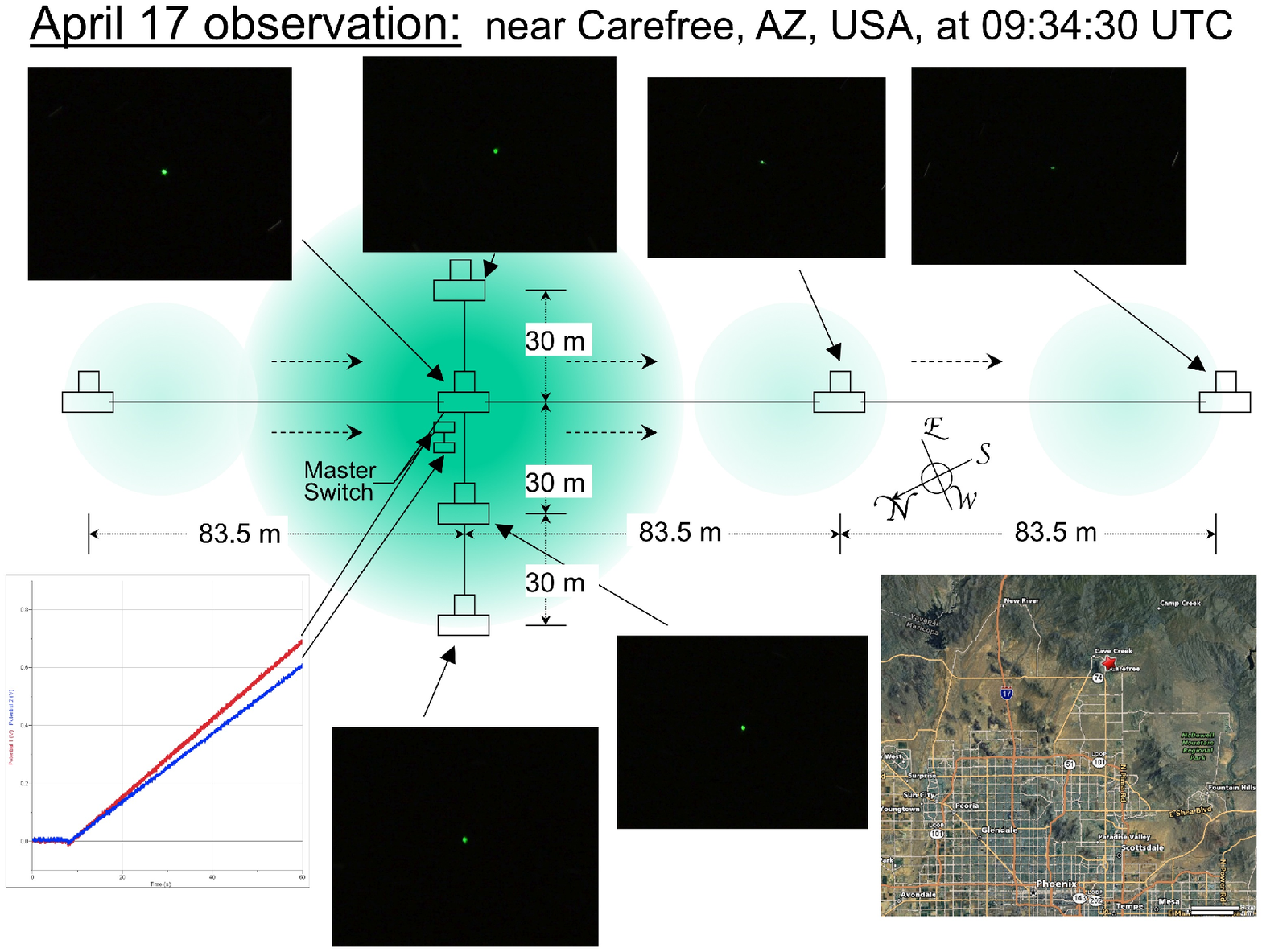}
\hfill
\includegraphics[angle=0,width=6.5cm]{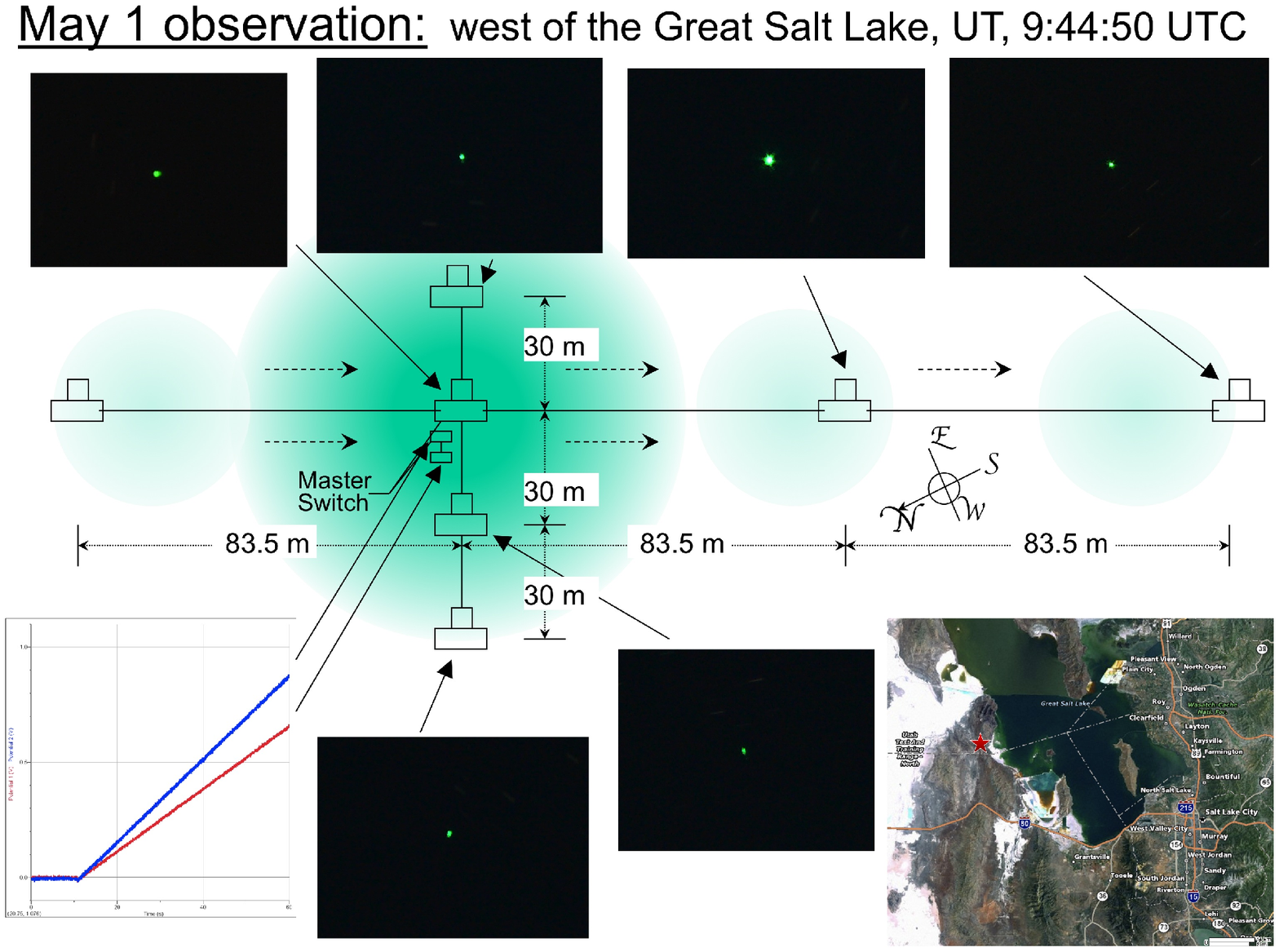}
\caption{
Seven-camera observations of CALIPSO overpasses, taken (left) on Apr.~17, 2007 near Carefree, Arizona, and (right) on May 1, 2007 near
the Great Salt Lake, Utah.
}
\label{fig:Carefreeobs}

\vspace*{9mm}

\includegraphics[angle=0,width=6.3cm]{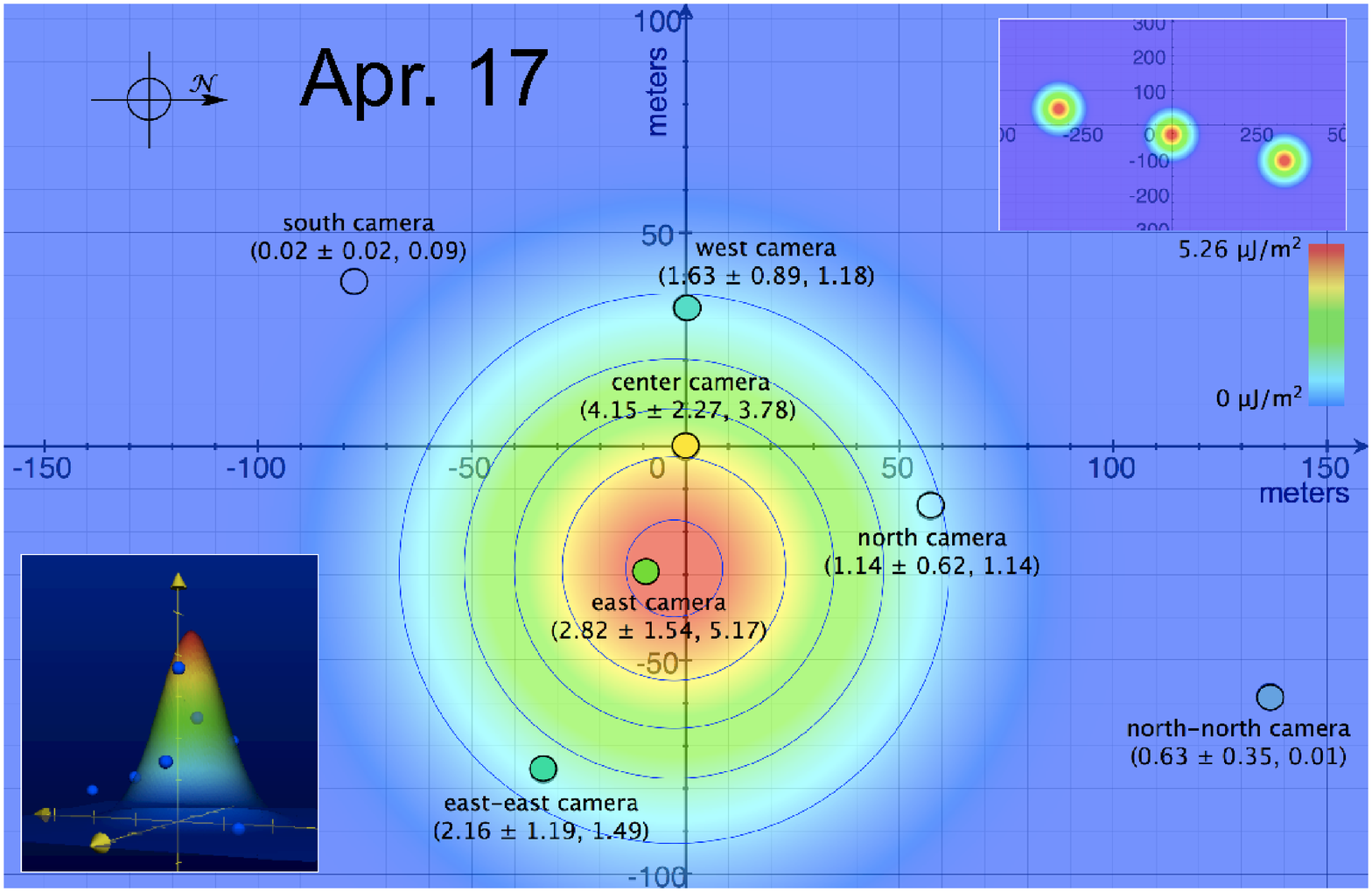}
\hfill
\includegraphics[angle=0,width=6.7cm]{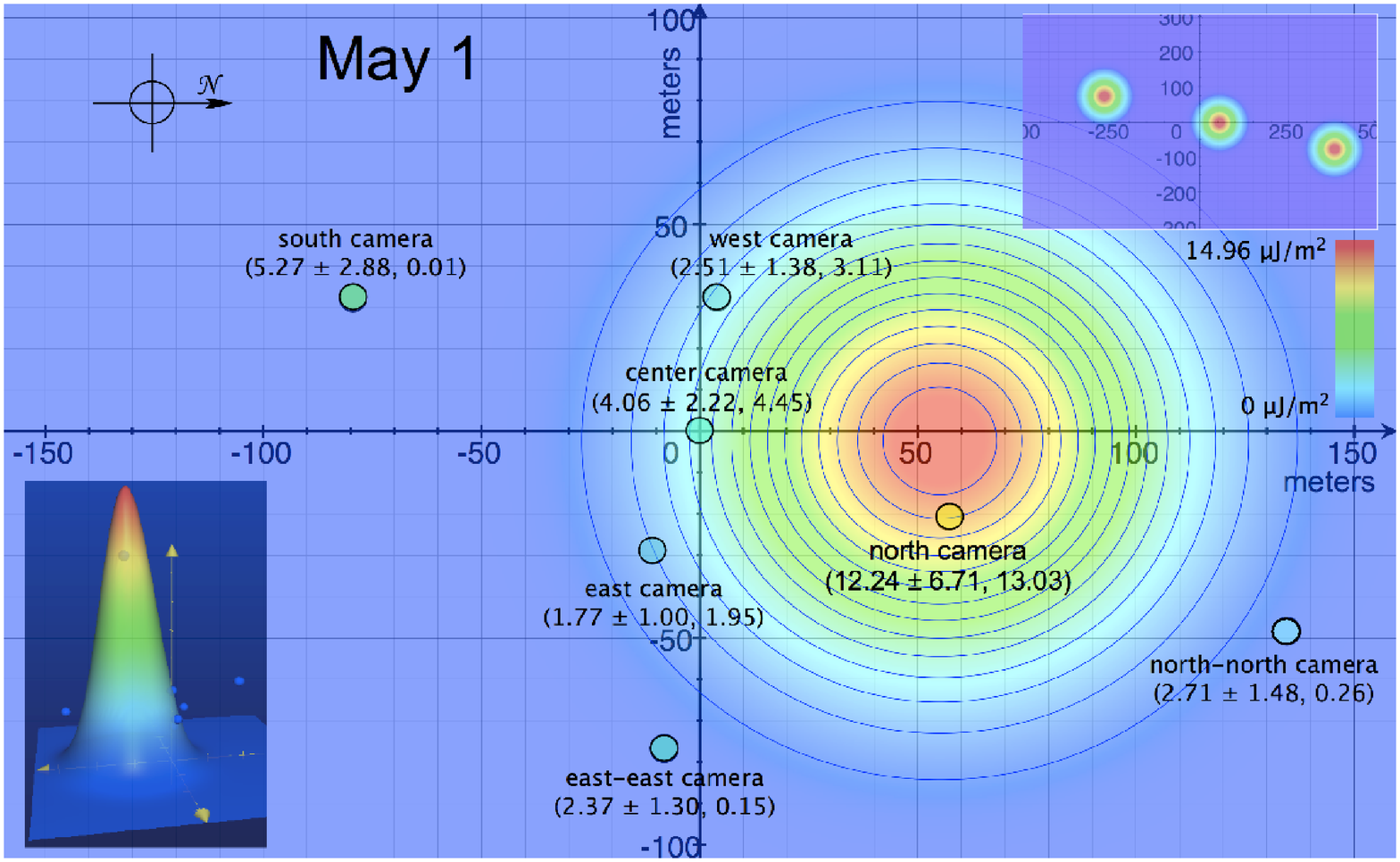}

\vspace*{1mm}

\includegraphics[angle=0,width=6.3cm]{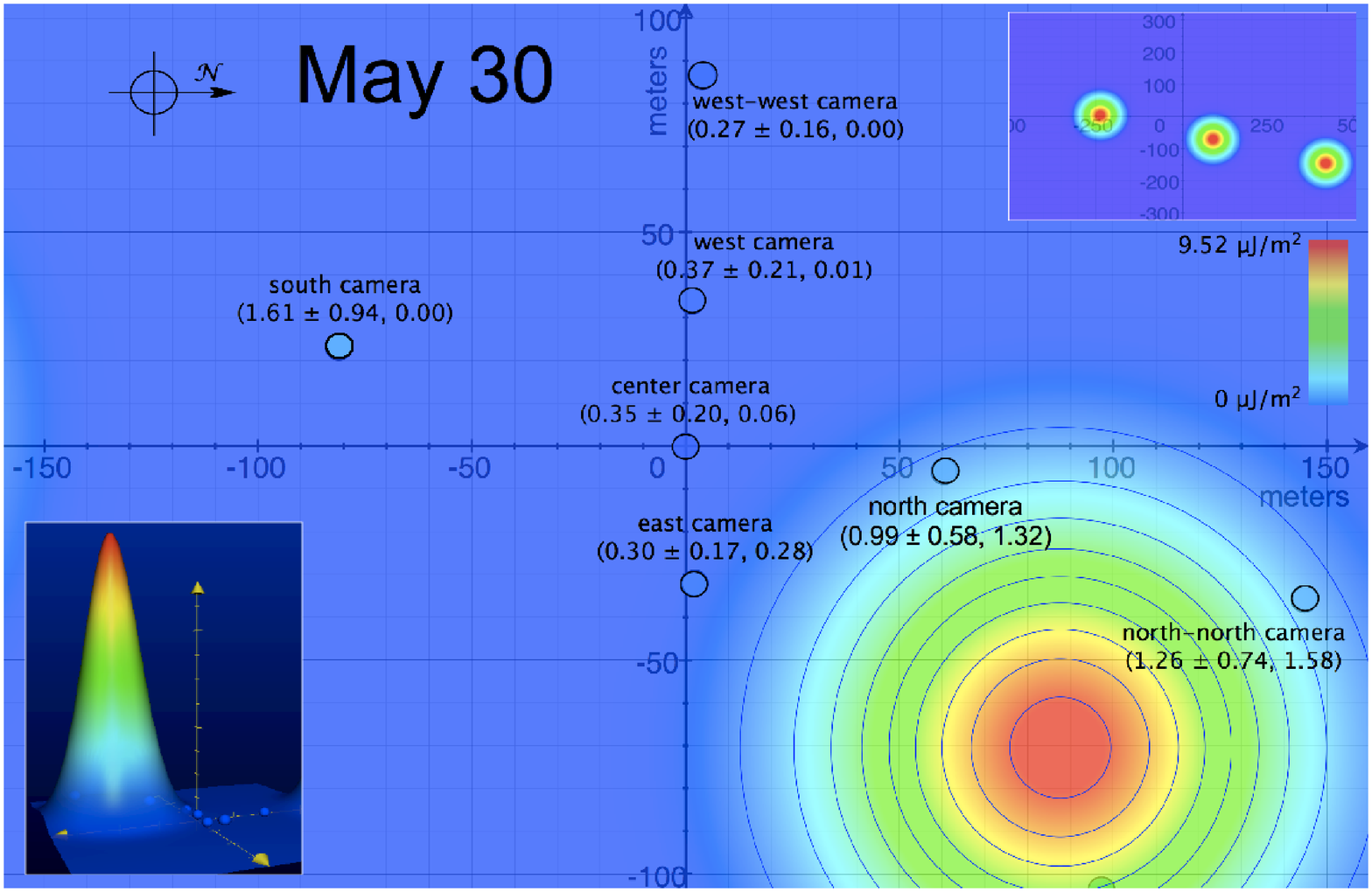}
\hfill
\includegraphics[angle=0,width=6.7cm]{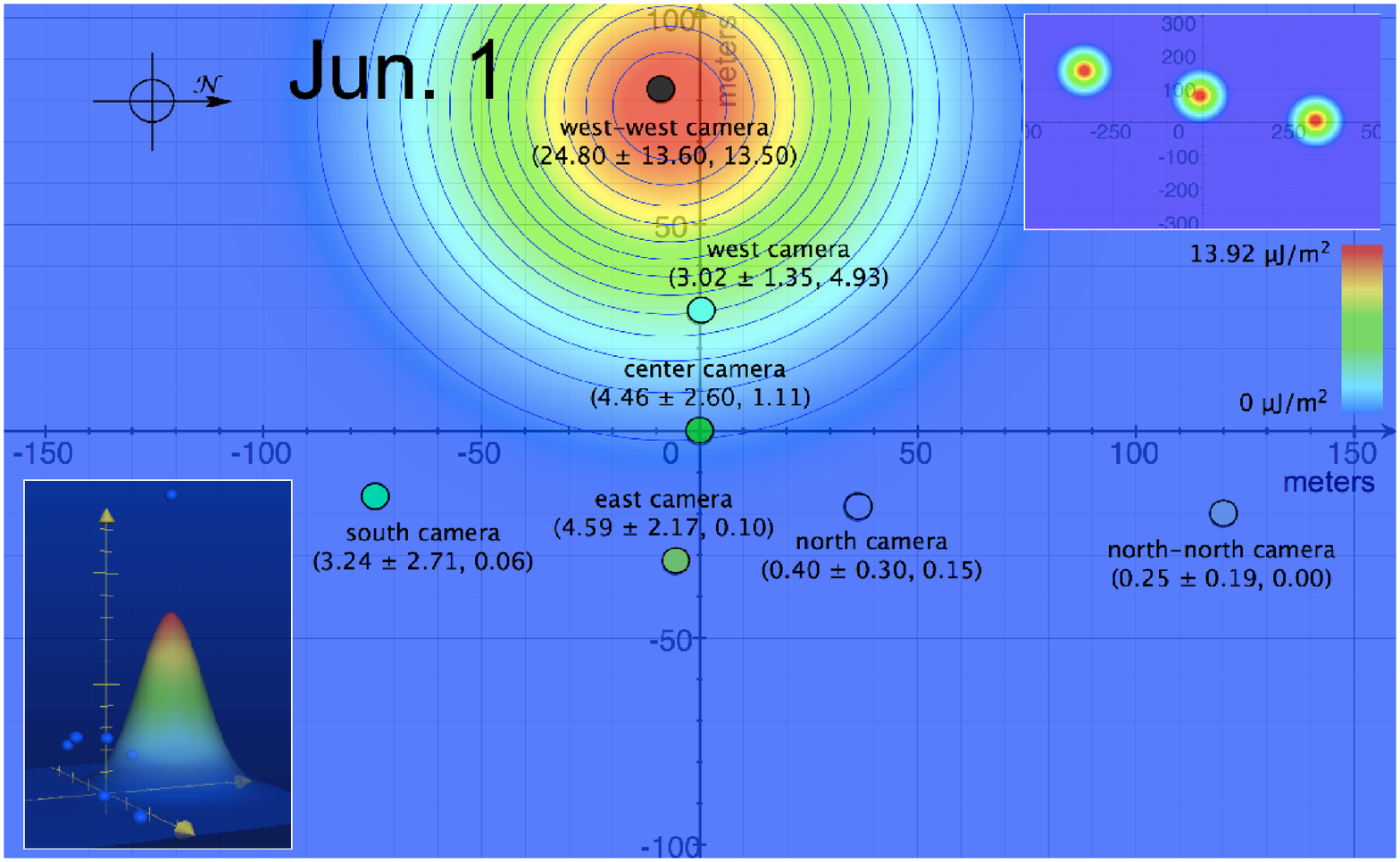}
\caption{
Camera time-integrated irradiance data, and the resulting fitted time-integrated irradiance maps, for the (upper left) Apr.~17, 
(upper right) May 1, (lower left) May 30, and (lower right) Jun.~1
observations.  The numbers at each camera refer to the measured value of time-integrated irradiance by that camera, with its associated
uncertainty (68\% CL), and the expectation value from the fitted function at the location of that camera.
The contours on each plot are spaced at 1 $\mu$J/m$^2$ intervals.
The upper-right inset on each plot extends the $x$ and $y$ axes in order to see the three 2-D Gaussians that comprise the 
fitted function (as described in the text), and the lower-left inset shows a different 3-D view (from the side, rather than from above) 
of the fitted function and the data points.
}
\label{fig:irradmaps}
\end{center}
\end{figure*}

\section{CALIPSO Data and Analysis}
\label{analysis}

For calibration of telescope optics and detector characteristics, authors \citep{stu06} have both conceived of and used a
wavelength-tunable laser within present and upcoming telescope domes as a color calibration standard.
Further progress and details can be found within the proceedings from Stubbs \& Tonry from this conference.
Although a wavelength-tunable laser calibration source in orbit \citep{alb06j} does not exist yet, at present there is a
532 nm laser in low-Earth orbit pointed toward the Earth's surface, with precise radiometric measurement of
the energy of each of the 20.25 Hz laser pulses, on the CALIPSO satellite, launched in April 2006 \citep{win09}.  We have
collected data from a portable network of seven cameras and two calibrated photodiodes, taken during CALIPSO flyovers
on clear days in various locations in western North America.  The cameras and photodiodes respectively capture images and
pulses from the eye-visible green laser spot at the zenith during the moment of a flyover.  Using precise pulse-by-pulse
radiometry data from the CALIPSO satellite, we compare the pulse energy received on the ground with the pulse energy
recorded by CALIPSO.  The ratio determines the atmospheric extinction.  

The CALIPSO (Cloud Aerosol Lidar and Infrared Pathfinder Satellite Observations) satellite was launched on April 28, 2006 as a 
joint NASA and CNES mission~\citep{win09}.  CALIPSO is part of a train of seven satellites (five of which are orbiting at the date 
of these proceedings), known as the ``A-Train,'' in sun-synchronous orbit at a mean altitude of approximately 690 km~\citep{sav08}. 
CALIPSO completes an orbit every 98.4 minutes 
(approximately 14.6 orbits per day), and repeats its track every 16 days.  CALIPSO contains a LIDAR (Light Detection and Ranging) 
system, known as CALIOP (Cloud Aerosol Lidar with Orthogonal Polarization), with a primary mission of obtaining high resolution 
vertical profiles of clouds and aerosols in the Earth's atmosphere~\citep{hun09}.  The CALIOP laser produces simultaneous, 
co-aligned 20 ns pulses of 532 nm and 1064 nm light, pointed a small angle (0.3$^\circ$) away from the geodetic nadir in the 
forward along-track direction, at a repetition rate of 20.16 Hz.  The light enters a beam expander, following which the divergence 
of each laser beam wavelength is approximately 100 $\mu$rad, producing a Gaussian spot of approximately 70 m RMS diameter on the ground.  
The pulse energy is monitored onboard the satellite, and averages approximately 110 mJ, at each one of the two wavelengths, per 
pulse.  The effective apparent magnitude of the 532 nm laser spot at the precise center of the beam is thus approximately -19.2,
however this high brightness, of course, falls off rapidly as one moves away from the center of the beam.

During 2007, the CALIPSO beam was observed at several locations in western North America using a portable ground station 
consisting of seven 
digital cameras, 
and two calibrated photodiodes co-located with the central camera.  
Two example observations can be seen in Fig.~\ref{fig:Carefreeobs}.
Observing locations were selected by means of the CALIPSO ground track, monitored by NASA.  

The cameras were calibrated to an absolute radiometric standard through the use of the NIST-calibrated Hamamatsu photodiode (see above)
and a low-noise 532 nm laser.
Measurements of the response linearity, an\-isotropy, and ambient temperature dependence of the cameras were also performed.  
Linearity was found to be maintained to $\pm2.4\%$.
The anisotropy 
of the response of the CCDs was 
measured by obtaining images as above with the laser spot at 900 different places on each of the cameras' CCDs.  The anisotropy, taken to
be the standard devation of those 900 measurements, was found to 
be 0.5\%.
Temperature calibration was performed using one of the cameras inside a sealed refrigerator
with a small hole in it to allow entrance of laser light (from inside a light-tight tube).  Over a temperature range from $20^{\circ}$C down to
$0^{\circ}$C, the image intensity was found to increase by the surprisingly large value of 52\%.

Knowing the precise relative positions of the cameras is necessary for extrapolating the camera energy measurements into a measurement of the 
total laser pulse energy at the ground.  During each observation, the location of each camera was measured using GPS, as well as a surveyor's tape measure to 
determine the relative positions via triangulation.  

CALIPSO measures the energy of each of its individual laser pulses.  The pulse energy monitoring on CALIPSO consists of NIST-calibrated 
photodiodes mounted on an integrating sphere, with 
pulse-by-pulse energy measurement with design absolute precision of $\pm2\%$ over full orbit, and relative precision of 
better than $\pm0.4\%$~\citep{win09}.  The pulse energies are recorded in the CALIPSO datasets available from the 
Atmospheric
Science Data Center (ASDC, located at NASA LaRC)\footnote{http://eosweb.larc.nasa.gov}, which are compared with the pulse energies measured at the ground to obtain
measurements of atmospheric extinction.  

Plots of the resulting measurements can be found in Fig.~\ref{fig:irradmaps}, and the measured values of atmospheric extinction, limited in uncertainty by atmospheric
extinction and the relatively poor knowledge of the shape of the laser beam, can be found in~\citep{alb12}.

\section{ALTAIR Balloon Program}
\label{balloon}

\begin{figure}[p]
\begin{center}
\includegraphics[angle=0,width=3.9cm]{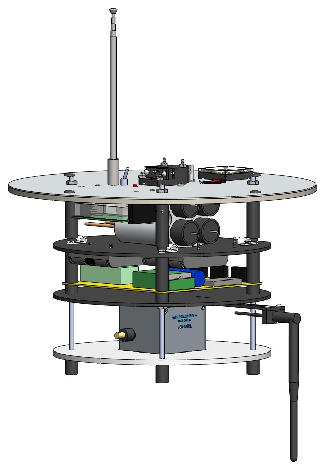}
\hspace*{4mm}
\includegraphics[angle=0,width=3.6cm]{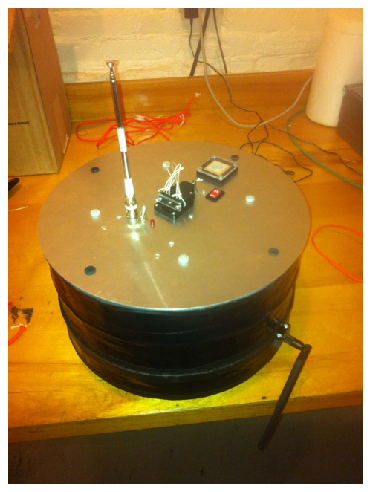}
\hspace*{4mm}
\includegraphics[angle=0,width=3.7cm]{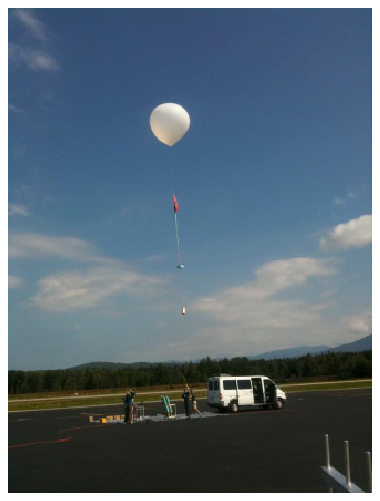}

\vspace*{1mm}

\includegraphics[angle=0,width=5.8cm]{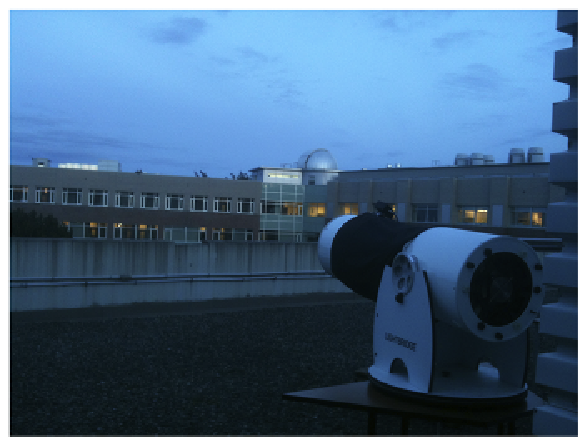}
\hspace*{6mm}
\includegraphics[angle=0,width=5.6cm]{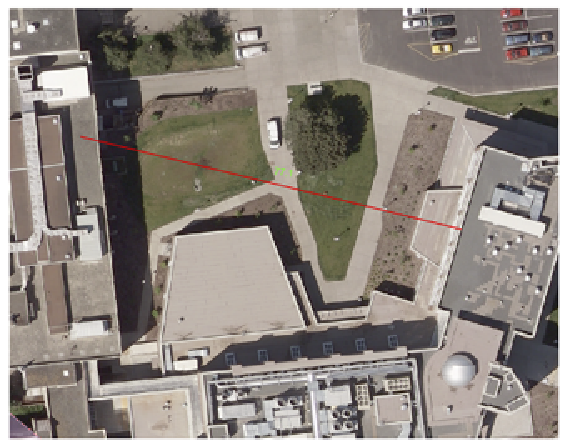}

\vspace*{1mm}

\includegraphics[angle=0,width=5.2cm]{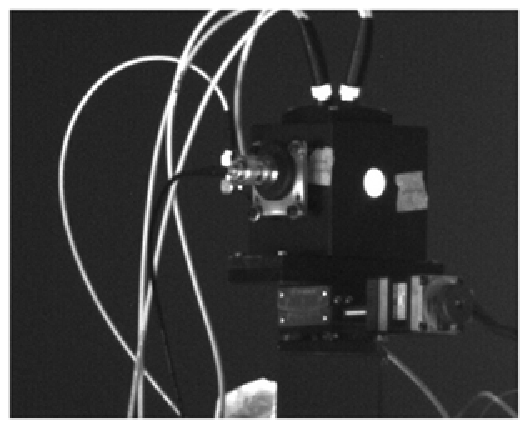}
\hspace*{5mm}
\includegraphics[angle=0,width=6.4cm]{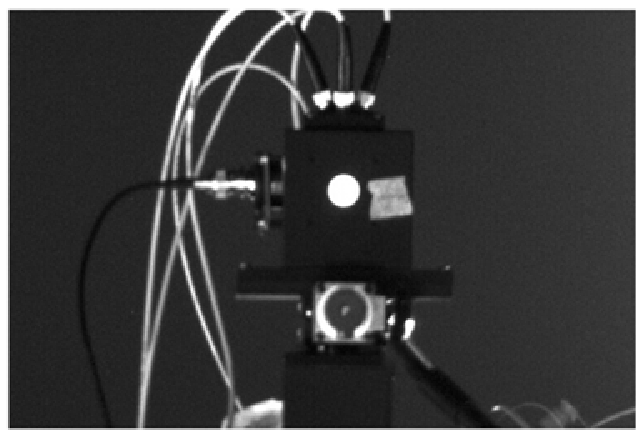}

\vspace*{1mm}

\includegraphics[angle=0,width=5.3cm]{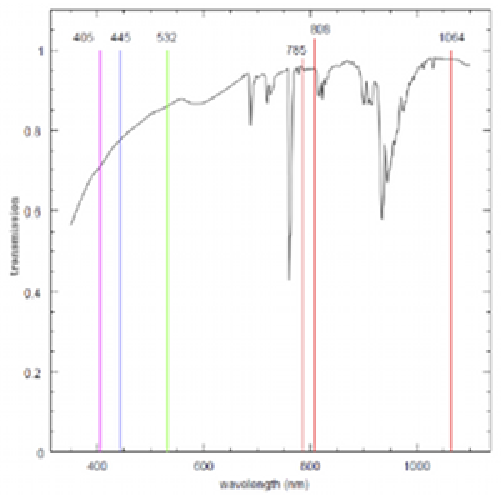}
\hspace*{7mm}
\includegraphics[angle=0,width=6.7cm]{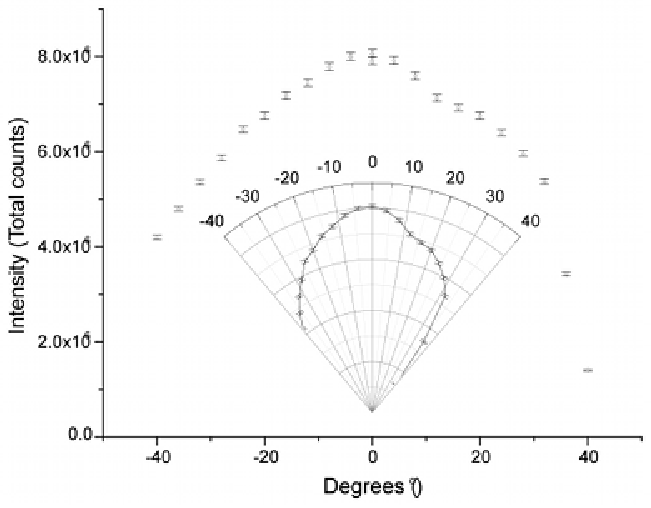}
\caption{
(Top row) 3-D solid diagram (left), photograph (center), and launch (right) of the balloon payloads.
(Second row) Setup for near-field imaging of the optical source: telescope (left) and view from above (right). 
(Third row) Near-field images of the optical source.  (Bottom row) Commonly-available laser lines
superimposed on a plot of average atmospheric transmission (left) and
measured light output distribution of the source (right).
}
\label{fig:PayloadSketch}
\end{center}
\end{figure}

Beam-based pulsed sources are certainly not the only options for photometric calibration standards above the atmosphere.  More isotropic,
and continuous, sources have the advantage that measurements are far less sensitive to the precise relative angle of the source and the observer,
and that one can time-average over atmospheric scintillation, rather than having to aperture-average.  Thus, a more isotropic calibrated source 
above the atmosphere is a more attractive option.  No such sources presently exist.  In order to test and fly such equipment, we have begun a
program, denoted ALTAIR (Airborne Laser for Telescopic Atmospheric Interference Reduction), 
of flights and observations of a small high-altitude balloon calibrated source platform, shown in Fig.~\ref{fig:PayloadSketch}.
These balloon flights will take place over major surveys, such as Pan-STARRS and LSST, in 2014 and beyond.
Initial flights in New Hampshire have begun, and flights over
Mt.~Hopkins, Arizona will begin later this year.
A balloon platform will not only provide critical data for precisely calibrating SNIa surveys, and also testing future satellite equipment, 
but additionally has the permanent advantage over satellite sources
that the payload may be recovered following the flight, and tested in the laboratory, rather than only having such laboratory tests preceeding the launch.
However, balloons of course cannot attain the altitude, nor the global reach, of a satellite.


Our initial payload light source contains laser diode module sources at 440, 532, 639, and 690 nm, directed into a 2'' diameter integrated sphere
via fiber optics, and monitored with a NIST-calibrated Hamamatsu S2281 photodiode.
In close parallel with our program of test flights, we have also begun a campaign of near-field observations of the source, via placement of the source
on a motorized mount on a rooftop, and observation with a telescope on a nearby rooftop approximately 75 m away, as shown in Fig.~\ref{fig:PayloadSketch}.
The addition of this man-made calibrated light source in near-space to the arsenal of techniques for 
photometric calibration will provide a powerful new tool for increasing precision in astrophysics.

\section{CONCLUSION}
\label{conclusion}

We have performed initial tests and measurements of space- and near-space-based precision photometric calibration sources.  Our initial tests are
highly promising, and we believe this is a viable means to reduce the dominant uncertainty in measurements of dark energy this decade.  Our
technique can additionally be applied to microwave astrophysics and to other regions of the spectrum, to impact cosmological and other
measurements in those areas as well.
Improved precision in photometric calibration will be nearly as critical for astronomy as increased aperture and etendue
telescopes in upcoming decades.  The
future of precision photometry is extremely promising, and laboratory-based standards in near-space and in space allow one to forsee
many-fold improvement in photometric calibration as a near-term prospect.

\acknowledgments

The authors would like to acknowledge the critical help and advice from Dr.~Matt Dobbs and Dr.~Keith Vanderlinde at McGill University,
Dr.~Susana Deustua at the Space Telescope Science Institute, and Dr.~Chris Pritchet and Mr.~Kevin Hildebrand at the University of Victoria. 
We would also like to acknowledge the work of the CALIPSO satellite team, led by PI Dr.~David M.~Winker and Project Scientist Dr.~Charles R.~Trepte
at NASA Langley Research Center (LaRC), and supported by NASA and CNES, as well as CALIPSO data obtained from the LaRC Atmospheric Science Data Center.
JEA is supported by Canadian Space Agency FAST grant \#11STFAVI41, Natural Sciences and Engineering Research Council of Canada grant
\#341310-2012, Canada Foundation for Innovation / British Columbia Knowledge and Development Fund grant \#13075, and the University of Victoria.

\bibliography{albert_j}

\end{document}